%% file: draft.tex
\documentclass[twocolumn,preprintnumbers,nofootinbib]{revtex4}
\usepackage{graphicx}% Include figure files
\usepackage{dcolumn}% Align table columns on decimal point
\usepackage{bm}% bold math
\usepackage{amsmath,amssymb,amsfonts}
\usepackage{latexsym}
\usepackage{bbm}
\usepackage{url} 
\usepackage{color}
\usepackage{soul}

\newcommand{\gev}{~\text{GeV}}
\def\nn{\nonumber}
\newcommand{\pb}{~\text{pb}}

\begin{document}
\preprint{ACFI-T19-01}

\title{CP-violating Dark Photon Interaction}

\author{Kaori Fuyuto$^1$}
\email{kfuyuto@umass.edu}
\author{Xiao-Gang He$^{2,3,4}$}
\email{hexg@phys.ntu.edu.tw}
\author{Gang Li$^{2}$}
\email{gangli@phys.ntu.edu.tw}
\author{Michael Ramsey-Musolf$^{1,5}$}
\email{mjrm@physics.umass.edu}
\affiliation{$^1$Amherst Center for Fundamental Interactions, Department of Physics,
University of Massachusetts Amherst, MA 01003, USA}
\affiliation{$^2$Department of Physics, National Taiwan University, Taipei 106}
\affiliation{$^3$T-D. Lee Institute and  School of Physics and Astronomy, Shanghai Jiao Tong University, 800 Dongchuan Road, Shanghai 200240}
\affiliation{$^4$National Center for Theoretical Sciences, Hsinchu 300}
\affiliation{$^5$Kellogg Radiation Laboratory, California Institute of Technology, Pasadena, CA 91125 USA}

\bigskip

\date{\today}

\begin{abstract}
We introduce a scenario for CP-violating (CPV)  dark photon interactions in the context of non-abelian kinetic mixing. Assuming an effective field theory that extends the Standard Model (SM) field content with an additional $U(1)$   gauge boson ($X$) and a $SU(2)_L$ triplet scalar, we show that there exist both CP-conserving and CPV dimension five operators involving these new degrees of freedom and the SM $SU(2)_L$ gauge bosons. The former yields kinetic mixing between the $X$ and the neutral $SU(2)_L$ gauge boson (yielding the dark photon), while the latter induces CPV interactions of the dark photon with the SM particles. 
We discuss experimental probes of these interactions using searches for permanent electric dipole moments (EDMs) and di-jet correlations in high-energy $pp$ collisions.  It is found that the experimental limit on the electron EDM currently gives the strongest restriction on the CPV interaction. In principle, high energy $pp$ collisions provide a complementary probe through azimuthal angular correlations of the two forward tagging jets in vector boson fusion. In practice, observation of the associated CPV asymmetry is likely to be challenging.

\end{abstract}

\maketitle

%-----------------------------------------------------------------------------------------------------------------------------------
%	Introduction
%-----------------------------------------------------------------------------------------------------------------------------------
\section{Introduction}
In recent years, the possible existence of a new $U(1)$ gauge boson has been motivated by experimental results in several phenomenological contexts, such as lepton flavor universalities in $B$ physics \cite{Sierra:2015fma, Allanach:2015gkd, Altmannshofer:2016jzy, Datta:2017pfz, Fuyuto:2017sys} and muon anomalous magnetic moment \cite{ Allanach:2015gkd, Harigaya:2013twa, Altmannshofer:2014cfa}. Moreover, the new gauge boson itself can be a dark matter candidate or mediator between the Standard Model (SM) particles and dark matter \cite{Altmannshofer:2016jzy, Tulin:2017ara, Battaglieri:2017aum}. It is often called a dark photon or $Z^\prime$.

The dark photon $(X)$ has kinetic mixing with the SM $U(1)_Y$ gauge boson, $ X_{\mu\nu} B^{\mu\nu}$, yielding interactions with the SM particles~\cite{Holdom:1985ag, Foot:1991kb} parameterized by a dimensionless parameter $\epsilon$. So far, a variety of searches for the $X$ in both low- and high-energy frontiers have been conducted~\cite{Essig:2013lka,Alexander:2016aln,Curtin:2014cca}. The resulting constraint is given by $\epsilon \lesssim 10^{-3}$, or smaller, for the dark photon mass below about 10~GeV.  
This situation motivates us to study theoretical explanations of the rather small coupling $\epsilon$. One solution might be going beyond renormalization theory, namely, introducing higher-dimensional operators \cite{Chen:2009dm, Chen:2009ab, Baek:2013dwa, Barello:2015bhq, Arguelles:2016ney}. 
In Ref. \cite{Arguelles:2016ney}, it is assumed that an $SU(2)_L$ triplet scalar $\Sigma$ $\sim (1,3,0)$, as well as the SM particles, are present below a scale $\Lambda$ that lies well above electroweak scale ($v\simeq 246~$GeV). This setup yields a $SU(2)_L$-invariant dimension-5 operator Tr$[W_{\mu\nu}\Sigma] X^{\mu\nu}/\Lambda$. After the triplet scalar develops a non-zero vacuum expectation value (VEV) $x_0$, the operator gives rise to kinetic mixing, $({x_0}/{\Lambda})  W^3_{\mu\nu} X^{\mu\nu}$. The triplet VEV breaks the custodial symmetry~\cite{Ross:1975fq}; therefore, it is strongly constrained by electroweak precision measurements~\cite{Tanabashi:2018oca}. As a result, kinetic mixing is naturally suppressed meanwhile collider signatures can be significant \cite{Arguelles:2016ney}. 

Most studies of the dark photon have concentrated on kinetic mixing that preserves CP symmetry. One may ask about the possibility of CP-violating (CPV) kinetic mixing, $B_{\mu\nu} \tilde X^{\mu\nu}$ with $\tilde X^{\mu\nu} = \epsilon^{\mu\nu\alpha\beta}X_{\alpha\beta}/2$. However, it is not present since the interaction is equivlant to total derivative and does not contribute to the action. Nevertheless, the current framework allows a CPV dimension-5 operator Tr$(W_{\mu\nu}\Sigma) \tilde X^{\mu\nu}/\Lambda$. Due to the presence of the triplet scalar, this operator is not equivalent to a total derivative.
Consequently, there are non-trivial CPV interactions that cannot be removed from the effective Lagrangian:  $(x_0+\Sigma^0)\tilde X^{\mu\nu}W^+_{\mu}W^-_{\nu}$ and $\tilde X^{\mu\nu}F_{\mu\nu}\Sigma^0$ with $\Sigma^0$ and $F_{\mu\nu}$ being a neutral component of the triplet scalar and field strength of the photon. The CPV interactions are a distinctive characteristic of non-abelian kinetic mixing, requiring different experimental probes from those of the CP-conserving (CPC) case.

Among powerful probes of CPV interactions at low energy are searches for permanent electric dipole moments (EDMs). The EDMs are P- and T-violating quantities, implying CP violation under the CPT theorem. They provide a powerful window on either strong CPV or CPV interactions arising from physics beyond the SM (BSM), since the predictions associated with SM electroweak CP violation are much below experimental sensitivities (for recent reviews, see Refs. \cite{Pospelov:2005pr, Engel:2013lsa, Yamanaka:2017mef, Chupp:2017rkp}). On the other hand,  BSM scenarios may possess new CPV phases, inducing nonzero EDMs that the experiments are able to reach.
In the present model, the CPV interaction $\tilde X^{\mu\nu}F_{\mu\nu}\Sigma^0$ generates the fermion EDMs at 1-loop level through two types of  mixing: non-abelian kinetic mixing and mixing of the SM Higgs boson with the neutral component of $\Sigma$. 
Although the experimental constraints on kinetic mixing generally become more severe as the dark photon is lighter, it is a notable that in this model a new  light degree of freedom, which cannot be integrated out, contributes to the EDMs. This situation contrasts with the more widely-considered sources of EDMs that involve new particles at the TeV scale and above.

Experiments at the high-energy frontier have the potential to play a complementary role in probing CPV interactions. In collider experiments, CPV effects can appear in angular distributions of final states. One possible way is to see a correlation of azimuthal angle difference between two tagging jets ($j$) in the  vector boson fusion (VBF) process \cite{Hagiwara:2009wt,Plehn:2001nj, Hankele:2006ma}. The aforementioned CPV interaction $\tilde X^{\mu\nu}W^+_{\mu}W^-_{\nu}$ can affect the angular correlation. Contrary to the EDMs, this collider signature does not depend on mixing beween the neutral scalars. Thus, we expect that the collider signature of the CPV interactions is potentially observable, having no suppression associated with a small scalar mixing term.

In this Letter, we will illustrate how the fermion EDMs probe the CPV dimension-5 operator and discuss the possibility of the complementary probe at the Large Hadron Collider (LHC). This paper is organized as follows. First, we introduce the dimension-5 operators and scalar potential. In Sec. III, it is discussed how the fermion EDMs arise from the CPV interactions, and their current bounds are shown. In Sec. IV, collider analyses associated with VBF processes are discussed. Section V contains our conclusions.

\label{sec:intro}

%-----------------------------------------------------------------------------------------------------------------------------------
%	Model
%-----------------------------------------------------------------------------------------------------------------------------------
\section{Model}
\label{sec:model}
The dimension-5 operators of interest  are 
\begin{align}
{\cal L}^{(d=5)}=-\frac{\beta}{\Lambda}{\rm Tr}\left[W_{\mu\nu}\Sigma \right]X^{\mu\nu}-\frac{\tilde \beta}{\Lambda}{\rm Tr}\left[W_{\mu\nu}\Sigma \right]\tilde X^{\mu\nu}, \label{dim5}
\end{align}
where $W_{\mu\nu}=W^a_{\mu\nu}\tau^a/2$ and $\tilde X^{\mu\nu}=\epsilon^{\mu\nu\alpha\beta}X_{\alpha\beta}/2$. An $SU(2)_L$ triplet scalar $\Sigma\sim(1,3,0)$ is given by
\begin{align}
\Sigma=\frac{1}{2}
\begin{pmatrix}
\Sigma^0 & \sqrt{2}\Sigma^+\\
\sqrt{2}\Sigma^- & -\Sigma^0
\end{pmatrix}.
\end{align} 
After the triplet scalar has its VEV $\langle \Sigma^0 \rangle = x_0$, the operators in Eq.~(\ref{dim5}) give the following interactions:
\begin{align}
&{\cal L}^{(d=5)}\supset -\frac{1}{2}\left(\alpha_{ZX}Z_{\mu\nu}X^{\mu\nu}+\alpha_{AX}F_{\mu\nu}X^{\mu\nu} \right)\nonumber\\
&-\frac{\tilde \beta}{2\Lambda}\tilde X^{\mu\nu}\left[s_WF_{\mu\nu}\Sigma^0
-ig_2(x_0+\Sigma^0)\left(W^-_{\mu}W^+_{\nu}-W^+_{\mu}W^-_{\nu}\right) \right], \label{CPV_int}
\end{align}
where $\alpha_{ZX(AX)}=\beta x_0c_W(s_W)/\Lambda$ with the weak mixing angle $c_W(s_W)\equiv\cos\theta_W(\sin\theta_W)$. $Z_{\mu\nu}$ and $F_{\mu\nu}$ are the field strengths of the $Z$ boson and photon, respectively.
The first row in Eq.~\eqref{CPV_int} comes from the CPC operator in Eq.~\eqref{dim5}, which implies kinetic mixing between the SM gauge bosons and dark photon. Taking $x_0 = 1$ GeV and $\Lambda=1$ TeV, one can see that the dimensionless kinetic mixing parameters are order of $10^{-3}$ for $\beta\sim \mathcal{O}(1)$. The second row describes the CPV interactions relevant to our study. While the first term of $\tilde X^{\mu\nu}F_{\mu\nu}\Sigma^0$ is responsible for the fermion EDMs, the subsequent terms contribute to VBF processes.
As mentioned in the Introduction, the interactions $\tilde X^{\mu\nu}A_{\mu\nu}$ and  $\tilde X^{\mu\nu}Z_{\mu\nu}$ are not present since each can be written as a total derivative. 

The scalar potential for the $SU(2)_L$ doublet and triplet scalars is~\cite{FileviezPerez:2008bj}
\begin{align}
V(H, \Sigma) =& -\mu^2H^{\dagger}H+\lambda(H^{\dagger}H)^2-\frac{M^2_{\Sigma}}{2}F+\frac{b_4}{4}F^2\nonumber\\
&+a_1H^{\dagger}\Sigma H+\frac{a_2}{2}H^{\dagger}HF,
\end{align}
where $F=(\Sigma^0)^2+2\Sigma^+\Sigma^-$ and $H=(\phi^+,~(h+i\phi^0)/\sqrt{2})$.
The last two terms with $a_1$ and $a_2$ cause mixing between $H$ and $\Sigma$. For the neutral scalars, we define their mass eigenstates as
\begin{align}
\begin{pmatrix}
H_1\\
H_2
\end{pmatrix}=
\begin{pmatrix}
\cos\theta & \sin\theta\\
-\sin\theta & \cos\theta
\end{pmatrix}
\begin{pmatrix}
h\\
\Sigma^0
\end{pmatrix},
\end{align}
with the mixing angle $\theta$ given by
\begin{align}
\tan2\theta = \frac{(-a_1+2a_2x_0)v}{2\lambda v^2-\left(2b_4x^2_0+\frac{a_1v^2}{4x_0}\right)}.
\end{align}
Here, $H_1$ is regarded as the SM Higgs with $m_{H_1}=125~$GeV. The above mixing allows the triplet scalar to couple to the SM fermions. For detailed expressions of the mass matrices, see Ref.~\cite{FileviezPerez:2008bj}. 

Besides the operators in Eq.~\eqref{dim5}, gauge invariance allows other operators at dimension $d\leq 5$: $B_{\mu\nu}X^{\mu\nu},~{\rm Tr}[W_{\mu\nu}\Sigma] B^{\mu\nu}$ and ${\rm Tr}[W_{\mu\nu}\Sigma] \tilde B^{\mu\nu}$. The first two operators can contribute to kinetic mixing,\footnote{The operator ${\rm Tr}[W_{\mu\nu}\Sigma] B^{\mu\nu}$ receives strong constraints from bounds on the $S$ parameter \cite{Tanabashi:2018oca}.} and the latter is able to give the CPV interaction. Here, in order to illustrate how the CPV observables are caused by the CPV dimension-5 interactions including the dark photon, we exclusively focus on the operators listed in Eq.~\eqref{dim5}. This setup can be realized if heavy degrees of freedom that induce the higher dimensional operators are not charged under $U(1)_Y$ \cite{Arguelles:2016ney}. Keeping these considerations in mind, we will investigate the probe of the CPV interactions in Eq.~\eqref{CPV_int} with the fermion EDMs and collider experiments.
%-----------------------------------------------------------------------------------------------------------------------------------
%	EDM
%-----------------------------------------------------------------------------------------------------------------------------------
\section{Electric Dipole Moment}
\label{sec:edm}
%--------------------------------------FIGURE----------------------------------
\begin{figure}[t]
\begin{center}
\includegraphics[width=4cm]{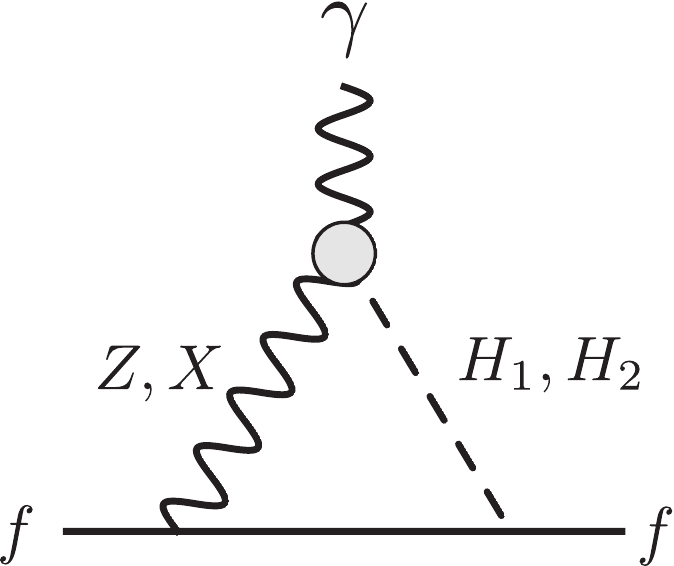} 
\end{center}
\caption{1-loop diagram of the fermion EDM generated by the CPV dimension-5 operator.}
\label{fig:edm}
\end{figure}
%-------------------------------------------------------------------------------------
Elementary fermion EDMs are defined by the interaction
\begin{align}
{\cal L}^{\rm EDM}=-\frac{i}{2}d_f\bar{f}\sigma^{\mu\nu}\gamma_5fF_{\mu\nu}.
\end{align}
One stringent limit on the CPV interactions comes from searches for the electron EDM, which have magnificently been improving the limit using polar molecules such as ThO and HfF$^+$. At $90\%$ confidence level (C.L.), the current upper limits are\footnote{The limits are obtained by assuming that only the electron EDM affects energy shifts of molecule systems. For recent discussions about exceptions to this assumption, see \cite{Chupp:2017rkp, Fleig:2018bsf, Fuyuto:2018scm}. }
\begin{align}
|d_e| &< 1.1 \times 10^{-29}~e~{\rm cm}~({\rm ThO} ~\mbox{\cite{Andreev:2018ayy}}),\\
|d_e| &< 1.3\times 10^{-28}~e~{\rm cm}~({\rm HfF^+}~ \mbox{\cite{Cairncross:2017fip}}).
\end{align}
The light quark EDMs constitute those of nucleons. The limit on the neutron EDM is given by \cite{Afach:2015sja}
\begin{align}
|d_n|<3.0\times 10^{-26}~e~{\rm cm},
\end{align}
at $90\%$ C.L.
The next generation EDM searches aim to improve the sensitivities by 
a factor of $10~(100)$ for $d_e~(d_n)$.
Moreover, the proton EDM experiment is also planned with usage of storage ring~\cite{Anastassopoulos:2015ura}. The prospective sensitivity is $|d_p|=1.0\times 10^{-29}~e~$cm.

In the present model, the third term in Eq.~\eqref{CPV_int} gives the CPV photon coupling to the dark photon and the neutral triplet scalar. The dark photon can couple to the SM fermions through kinetic mixing, whereas the neutral triplet scalar can be connected to them through mixing with the doublet scalar. It follows that the fermion EDMs arise at 1-loop level as in Fig. \ref{fig:edm}. \footnote{In Ref.\cite{Dall:2015bba}, another type of the dimension-5 operator is discussed for the fermion EDMs.} The resulting  fermion EDM is
\begin{align}
d_f = &\frac{e}{8\pi^2}\frac{m_f}{v}c_{\theta}s_{\theta}\bigg[C_ZV^f_Zf\left(r_{ZH_1},~r_{ZH_2}\right)\nonumber\\
&\hspace{2cm}+C_XV^f_Xf\left(r_{XH_1},~r_{XH_2}\right) \bigg], \label{fermion_edm}
\end{align}
where $r_{Z(X)H}=m^2_{Z(X)}/m^2_{H}$, and the loop function is
\begin{align}
f(x,y)=\frac{1}{2}\log{\left(\frac{m^2_{H_1}}{m^2_{H_2}} \right)}-\frac{1}{2}\left(\frac{x\log x}{1-x}-\frac{y \log y}{1-y}\right).
\end{align}
The couplings are given by
\begin{align}
C_Z&=\frac{\tilde \beta}{\Lambda}s_Ws_{\xi},\hspace{1.5cm}
C_X=\frac{\tilde \beta}{\Lambda}s_Wc_{\xi},\\
V^f_Z&=\left(c_{\xi}-s_{\xi}\alpha_{ZX} \right)\frac{g^f_Z}{c_Ws_W}-Q_f\alpha_{AX}s_{\xi}, \label{VfZ}\\
V^f_X&=-\left(s_{\xi}+c_{\xi}\alpha_{ZX} \right)\frac{g^f_Z}{c_Ws_W}-Q_f\alpha_{AX}c_{\xi}, \label{VfX}
\end{align} 
where $Q_f$ denotes the fermion electric charge and $g_Z^f=I/2-s_W^2Q_f$ with isospin charge $I$. The mixing angle $\xi$ is introduced to diagonalize the mass matrices of the SM $Z$ boson and dark photon:
\begin{align}
\tan2\xi = -\frac{2m^2_Z\alpha_{ZX}}{m^2_Z-m^2_X}.
\end{align}
Except near the region $m_Z\simeq m_X$, the mixing angle can be expressed by 
$\xi = -m_Z^2\alpha_{ZX}/(m^2_Z-m^2_X)$ since $\alpha_{ZX}\sim O(10^{-3})$.
Assuming that $s_{\xi} \sim \alpha_{ZX}$ for $m_Z \gg m_X$, we see that $d_f\propto \alpha_{ZX, AX}\tilde\beta/\Lambda$; therefore, the fermion EDMs  scale as $1/\Lambda^2$. Furthermore, as seen in Eq.~\eqref{fermion_edm}, the fermion EDMs decrease as $\sin\theta$ approaches zero. 
%--------------------------------------FIGURE----------------------------------
\begin{figure}[t]
\begin{center}
\includegraphics[width=7cm]{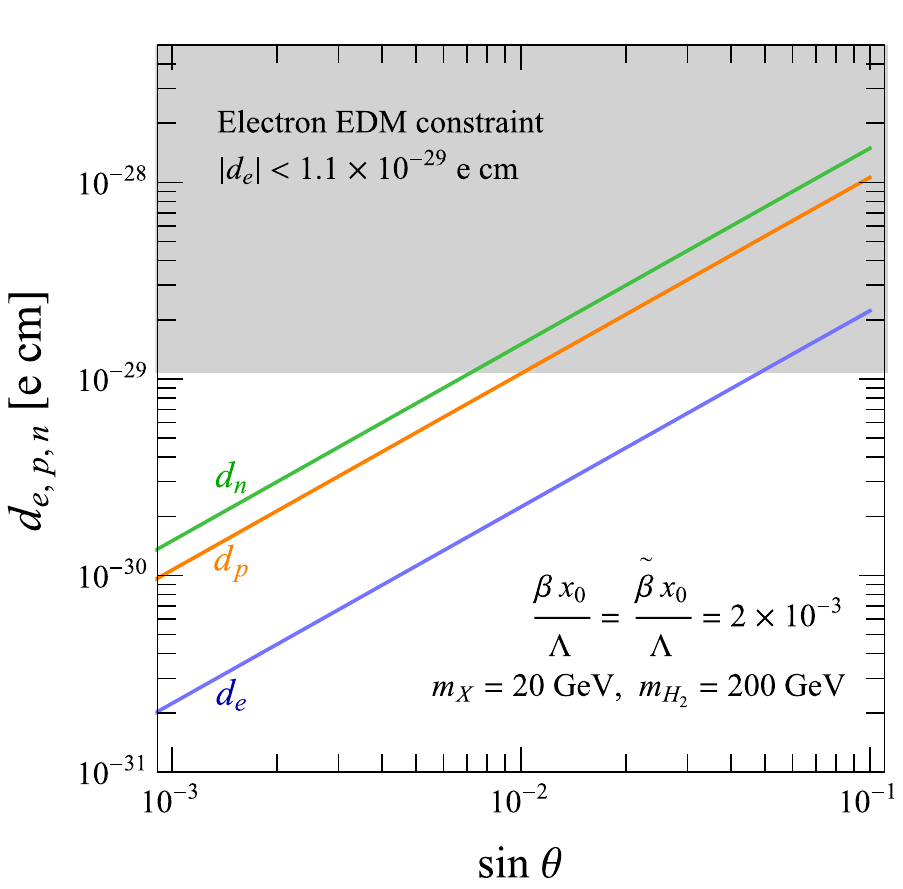} 
\end{center}
\caption{The electron, proton and neutron EDMs against the mixing parameter $\sin\theta$. It is taken that $\beta x_0\Lambda=\tilde\beta x_0\Lambda = 2 \times 10^{-3},~m_X=20~{\rm GeV}$ and $m_H=200~{\rm GeV}$.}
\label{fig:epn_edm}
\end{figure}
%-------------------------------------------------------------------------------------

Figure \ref{fig:epn_edm} shows predictions for the EDMs as a function of the mixing parameter $\sin\theta$. For an illustration of the EDMs induced by a light degree of freedom, we take the relatively light dark photon mass $m_X=20$ GeV. It should be emphasized that the values of the EDMs do not drastically change  for even lighter $m_X$, and in principle a non-vanishing EDM can arise from the exchange of an ultralight ($m_X$ at the MeV-scale and below) dark photon. However,
the CPC kinetic mixing angle is restricted more severely for the MeV-scale dark photon. Other relevant parameters are fixed at $\beta x_0/\Lambda =\tilde\beta x_0/\Lambda = 2\times 10^{-3}$ and $m_{H_2}=200~{\rm GeV}$.  At this benchmark point, the second term in Eq.~\eqref{VfX} becomes the leading one while it is clear that the dominant contribution to $V_Z^f$ comes from the first term in Eq. (\ref{VfZ}). The blue line represents the electron EDM, and the shaded region is the current experimental bound indicating $\sin\theta \lesssim 5\times 10^{-2}$.  The green and orange lines correspond to the neutron and proton EDMs, for which theoretical formulae obtained by the lattice QCD calculations are employed \cite{Bhattacharya:2015esa, Bhattacharya:2015wna}. Naively, they become larger by $m_q/m_e$ than $d_e$. In addition, since the neutron EDM receives the dominant contribution from the down-quark EDM, it somewhat exceeds $d_p$. The experimental bound on $d_n$ is not reflected in the current figure since it is located well above the chosen range. It is also seen that the prospective sensitivity for $d_p$ is able to reach $\sin\theta\sim10^{-2}$. 

It should be noted that the 2-loop contribution without scalar mixing is also present. The contribution is induced by the so-called Barr-Zee diagram~\cite{Barr:1990vd}, in which the $W$ boson runs in the upper loop. Naive dimensional analysis shows that $d_f^{2-{\rm loop}}/d_f^{1-{\rm loop}}\sim \frac{1}{(4\pi)^2} \frac{vx_0}{m^2_W} \frac{c_{\theta}}{s_{\theta}}\sim 10^{-4}\frac{c_{\theta}}{s_{\theta}}$, which implies that the 2-loop diagram can be comparable with the 1-loop contribution if $s_\theta\sim 10^{-4}$.  However, as expected from Fig. \ref{fig:epn_edm}, such a region indicates that the EDMs are below the prospective sensitivities at the next generation EDM experiments.  Therefore, it is sufficient to include only the 1-loop contribution in the current analysis.

%-----------------------------------------------------------------------------------------------------------------------------------
%	Collider
%-----------------------------------------------------------------------------------------------------------------------------------
\section{Collider Probes: Dijet Correlations}
\label{sec:collider}
The previous study of the CPC operator in Eq.~\eqref{dim5}  showed that for appropriate choice of final states involving two $X$ and one or more electroweak gauge bosons, the LHC production rate need not be suppressed by the mixing parameter $\epsilon\sim x_0/\Lambda$ \cite{Arguelles:2016ney}.  The corresponding collider phenomenology, thus, contrasts with one wherein the $X$ interacts via abelian kinetic mixing and/or mixing of a dark Higgs with the SM Higgs doublet. The CPC non-abelian kinetic mixing  yields a unique set of collider signatures that may be exploited for discovery.

Here, we explore the extent to which collider studies may also provide a complementary probe of the CPV operator. An interesting set of observables involves azimuthal angular correlations between the forward, \lq\lq tagging jets" $j$ produced in VBF process: $pp\to jj {\mathbf{X}}$, where ${\mathbf{X}}$ denotes other objects produced in the underlying hard event. Dijet azimuthal angle distributions that depend on $\cos\Delta\phi_{jj}$ have been considered as a means for discovering an invisibly-decaying Higgs boson \cite{Eboli:2000ze}, diagnosing the spin of new particles that appear in pairs in the state  ${\mathbf{X}}$ \cite{Buckley:2010jv, Buckley:2014fqa}, and searching for higher spin resonances \cite{Hagiwara:2009wt}. The interference between CPC and CPV interactions in the fusion vertex could lead to a $\sin\Delta\phi_{jj}$ dependence in the $\Delta\phi_{jj}$ distribution, a feature that has been considered as a means of determining the CP nature of the Higgs boson in VBF \cite{Plehn:2001nj, Hankele:2006ma}.

In the present context, the simplest VBF final state has ${\mathbf{X}}\to X$ as indicated by the interaction $(x_0/\Lambda) \tilde X^{\mu\nu}W_{\mu}^+W^-_{\nu}$ in Eq.~\eqref{CPV_int}.  The observable of interest in this case is the CPV asymmetry
\begin{align}
\label{eq:asymmetry}
\mathcal{A}=\dfrac{\sigma(\sin\Delta\phi_{jj}>0)-\sigma(\sin\Delta\phi_{jj}<0)}
{\sigma(\sin\Delta\phi_{jj}>0)+\sigma(\sin\Delta\phi_{jj}<0)}.
\end{align}
In the laboratory frame,  $\Delta\phi_{jj}$ is defined by \cite{Hankele:2006ma,  Englert:2012xt, Hagiwara:2009wt,Nakamura:2016agl}
\begin{align}
\Delta \phi_{jj}=\phi_{j_1}-\phi_{j_2},
\end{align}
where $\phi_{j_1}$ and $\phi_{j_2}$ are the azimuthal angles of the jets in the forward and backward regions of the detector, respectively.
$\sigma(\sin\Delta\phi_{jj}>0)$ and $\sigma(\sin\Delta\phi_{jj}<0)$ denote the total cross sections (signal plus background) for $0\leq\Delta\phi_{jj} \leq \pi$ and $-\pi\leq\Delta\phi_{jj} \leq 0$, respectively. 

The rate for this process is suppressed by $(x_0/\Lambda)^2$, leading to production cross sections of $\mathcal{O}(\mathrm{fb})$ or smaller for phenomenogically allowed parameter choices. Nevertheless, one may expect a sufficiently large number of events at the high-luminosity phase of the LHC with 3 ab$^{-1}$ of integrated luminosity. To proceed, we choose a representative choice of parameters consistent with the EDM-sensitive region: $\beta x_0/\Lambda=1\times 10^{-3}$ and $\tilde{\beta}x_0/\Lambda=4\times 10^{-3}$. We generate signal process $pp\to jjX$ using \texttt{MadGraph5\_aMC@NLO}~\cite{Alwall:2014hca} with the cuts:
\begin{align}
\label{eq:basic_cuts}
p_{T}^{j}>20\gev,\quad \Delta R_{jj}>0.4,\quad |y_j|<5,\nn\\
|\Delta y_{jj}|>4.2,\quad
y_1\cdot y_2<0,\quad m_{jj}>600\gev.
\end{align}
In the above, $j$ denotes light-flavor quarks, and the angular distance in the $\eta-\phi$ plane is defined as $\Delta R_{ij}\equiv\sqrt{(\eta_i-\eta_j)^2+(\phi_i-\phi_j)^2}$ with $\eta_i$ and $\phi_i$ being the pseudo-rapidity and azimuthal
angle of particle $i$, respectively. $p_T^j$, $y_j$ denote the transverse momentum and rapidity of jet $j$. $y_1$ and $y_2$ are the rapidities in the forward and backward regions of the detector. $\Delta y_{jj}$ and $m_{jj}$ are the rapidity difference and invariant mass of these two jets. The NN23LO1 Parton Distribution Function (PDF) set~\cite{Ball:2012cx} is used. 

We choose two benchmark values of the dark photon mass, $m_X=30,100\gev$. The signal cross sections for $pp\to jjX$ are $1.85\times 10^{-4}\pb$ and $2.51\times 10^{-3}\pb$ for $m_X=30\gev$ and $100\gev$, respectively, after imposing the cuts in Eq.~\eqref{eq:basic_cuts}. We note that these cross sections include non-VBF subprocesses, such as those wherein the $X$ is emitted from a quark line rather than fusing weak vector bosons. The corresponding asymmetries with zero SM backgrounds are $0.009$ and $0.021$. To enhance these asymmetries, we observe that the non-VBF $X$ production process tends to yield a softer $p_T^j$ and $p_T^X$ spectrum than does the VBF subprocesses. Thus we impose the additional cuts 
\begin{align}
p_{T}^{j}>40\gev,\quad p_{T}^{X}>70\gev
\end{align}
for $m_X=30\gev$ and
\begin{align}
p_{T}^{j}>60\gev,\quad p_{T}^{X}>100\gev
\end{align}
for $m_X=100\gev$.
As a result, the asymmetries are increased. We obtain $\mathcal{A}(m_X=30\gev) = 0.017$  and $\mathcal{A}(m_X=100\gev)=0.135$ with zero SM backgrounds. However, the respective cross sections $\sigma(pp\to jjX)$ are  reduced to $4.40\times 10^{-5}\pb$ and $1.68\times 10^{-4}\pb$.

To suppress the SM backgrounds, we consider the displaced decays of the $X$ to $\ell^+\ell^-$ pairs ($\ell = e, \mu$) with branching ratios $0.32$ and $0.07$ for $m_X=30\gev$ and $100\gev$, respectively. The resulting respective numbers of events are $42$ and $35$. It is clear that the associated statistical precision is, thus, not sufficient to permit observation of a CPV asymmetry in the $\mathcal{O}(1-10\%)$ range. 

A potentially more promising possibility involves an explicit triplet-like scalar in the final state, which stems from the interaction $\Sigma^0\tilde X^{\mu\nu}W_{\mu}^+W^-_{\nu}/\Lambda$ in Eq.~\eqref{CPV_int}. In this case one avoids the $(x_0/\Lambda)^2$ suppression factor, though with the price of an additional final state particle phase space. For concreteness, we consider the case ${\mathbf{X}} = X H_2$. From an analysis of previous long lived particle searches \cite{Aad:2014yea, ATLAS2016042}, we find that the choice $\beta/\Lambda$ $=\tilde{\beta}/\Lambda$ $=1/\text{TeV}$ is allowed. In this case, we find that after imposing the same selection cuts as in Eq.~(\ref{eq:basic_cuts}) and considering the displaced $X$ decays to di-lepton pairs, we would expect roughly 1500 signal events after collection of 3 ab$^{-1}$ of data. The corresponding statistical uncertainty is 2.6\% without the SM backgrounds. On the other hand, we find that the magnitude of CPV asymmetry $\mathcal{A}$ lies below one percent. While it may be possible to impose additional cuts to enhance the latter (as in the case of the $pp\to jjX$ study), it appears challenging to probe the CPV interactions from the CPV operator in Eq.~\eqref{dim5} through VBF process at the LHC.

%-----------------------------------------------------------------------------------------------------------------------------------
%	Conclusions
%-----------------------------------------------------------------------------------------------------------------------------------
\section{Conclusions}
\label{sec:conclusion}
The dark photon is a new $U(1)$ gauge boson, which is motivated by several phenomenological considerations. It couples to the SM fermions through kinetic mixing with the SM gauge bosons.  CPV kinetic mixing does not arise in the abelian mixing case since the operator $X_{\mu\nu} {\tilde B}^{\mu\nu}$ is equivalent to a total derivative. However, in the non-abelian mixing context, the dark photon can have CPV interactions. One interesting source of  non-abelian kinetic mixing is the higher dimensional operator Tr$[W_{\mu\nu}\Sigma]X^{\mu\nu}/\Lambda$, which can naturally explain the small mixing parameter. The corresponding CPV operator Tr$[W_{\mu\nu}\Sigma]\tilde X^{\mu\nu}/\Lambda$  becomes a source for the fermion EDMs, which are induced by the 1-loop diagram with the help of CPC kinetic and scalar mixing. Therefore, as far as the mixing parameters are nonvanishing, the CPV operator can be probed by searches for the EDMs of the electron, neutron and proton. A potentially complementary probe might be studies of the VBF process at the Large Hadron Collider that analyze azimuthal angular correlations of the two forward tagging jets. Importantly, this process is free from scalar mixing and, thus, unsuppressed by the small scalar mixing angle. Here, we have considered two possible VBF channels: $pp\to jj X$ and $pp\to jj XH_2$. We find that the former suffers from large statistical uncertainty, and the latter cannot produce a sufficiently large CPV asymmetry to be observed. Consequently, the EDM searches provide the most promising avenue for probing the CPV dark photon interaction.

\begin{acknowledgments}
We would like to thank Junya Nakamura for helpful discussions. XGH and GL are supported in part by the MOST (Grant No. MOST 106-2112-M-002-003-MY3 ), Key Laboratory for Particle Physics, Astrophysics and Cosmology, Ministry of Education, and Shanghai Key Laboratory for Particle Physics and Cosmology (Grant No. 15DZ2272100), and in part by the NSFC (Grant Nos. 11575111 and 11735010).
KF and MRM are supported in part under U.S. Department of Energy contract DE-SC0011095.
\end{acknowledgments}

%-----------------------------------------------------------------------------------------------------------------------------------
%-----------------------------------------------------------------------------------------------------------------------------------
\input bib
\end{document}

%% file: bib.tex
%Reference